\documentclass{aa}
\usepackage{graphicx}
\newcommand{\harps}{{\textsc{harps}}}
\newcommand{\sophie}{{\textsc{sophie}}}
\newcommand{\elodie}{{\textsc{elodie}}}
\newcommand{\cible}{{Procyon}}

\newcommand{\deuxzero}{\delta\nu_{02}}  

\newcommand{\nident}{28}  
\newcommand{\ind}[1]{_{\rm #1}}
\newcommand{\courb}{{\cal C}}
\newcommand{\diff}{{\rm d}}
\def\ms{\,m\,s$^{-1}$}         
\def\cms{\,cm\,s$^{-1}$}       
\def\m2s2{\,m$^{2}$\,s$^{-2}$} 

\begin{document}
\title{Asteroseismology of Procyon with  SOPHIE
\thanks{Based on observations collected
with the \sophie\ \'echelle spectrometer mounted on the 1.93-m telescope at OHP, France (program  06B.PNPS.BOU); {http://www.obs-hp.fr/www/guide/sophie/sophie-eng.html}
}
\thanks{Data corresponding to Table 1 and Fig 1 are only available in electronic form
at the CDS via anonymous ftp to cdsarc.u-strasbg.fr (130.79.128.5)
or via http://cdsweb.u-strasbg.fr/cgi-bin/qcat?J/A+A/}
}
\titlerunning{Procyon oscillations with SOPHIE}
\author{
B. Mosser\inst{1}\and
F. Bouchy\inst{2}\and
M. Marti\'c\inst{3}\and
T. Appourchaux\inst{4}\and
C. Barban\inst{1}\and
G. Berthomieu\inst{5}\and
R.A. Garcia\inst{6}\and
J.C. Lebrun\inst{3}\and
E. Michel\inst{1}\and
J. Provost\inst{5}\and
F. Th\'evenin\inst{5}\and
S. Turck-Chi\`eze\inst{6}
}
\institute{%
LESIA, CNRS, Universit\'e Pierre et Marie Curie, Universit\'e Denis Diderot, Observatoire de Paris, 92195 Meudon cedex, France\\
\email{benoit.mosser@obspm.fr}
\and
Institut d'Astrophysique de Paris, CNRS, Universit\'e Pierre et Marie Curie, 98$^{bis}$ bld Arago, 75014 Paris, France\\
\and
Service d'A\'eronomie du CNRS, BP 3, 91371 Verri\`eres le Buisson, France\\
\and
Institut d'Astrophysique Spatiale, Universit\'e Paris XI-CNRS, B\^atiment 121, 91405 Orsay cedex, France\\
\and
Laboratoire Cassiop\'ee, UMR CNRS 6202, Observatoire de la C\^ote d'Azur, BP 4229, 06304 Nice cedex 4, France\\
\and
DAPNIA/DSM/Service d'Astrophysique, CEA/Saclay, 91191 Gif-sur-Yvette Cedex, France\\
}
\date{Submitted July 23, 2007}

\abstract{This paper reports a 9-night asteroseismic observation program conducted in January 2007 with the new spectrometer \sophie\ at the OHP 193-cm telescope, on the F5 IV-V target Procyon A.}%
{This first asteroseismic program with \sophie\ was intended to test the performance of the instrument with a bright but demanding asteroseismic target and was part of a multisite network.}%
{The \sophie\ spectra have been reduced with the data reduction software provided by OHP. The Procyon asteroseismic data were then analyzed with statistical tools. The asymptotic analysis has been conducted considering possible curvature in the \'echelle diagram analysis.}%
{These observations have proven the efficient performance of \sophie\ used as an asteroseismometer, and succeed in a clear detection of the large spacing. An \'echelle diagram based on the 54-$\mu$Hz spacing shows clear ridges. Identification of the peaks exhibits large spacings varying from about 52 $\mu$Hz to 56~$\mu$Hz. Outside the frequency range [0.9, 1.0\,mHz] where the identification is confused,
the large spacing increases at a rate of about $\diff \Delta\nu/\diff n \simeq 0.2\,\mu$Hz. This may explain some of the different values of the large spacing obtained by previous observations.}
{}
\maketitle
\section{Introduction}

Asteroseismology consists in measuring properties of oscillation modes and provides a unique tool for ``drilling'' stellar interiors and for testing the stellar internal structure and evolution model. In recent years Doppler ground-based observations have provided detection and identification of p-modes in an increasing list of bright solar-like stars (\cite{2006ESASP.624E..25B}). New generation high-precision spectrometers such as \harps\ (\cite{2004A&A...423..385P}) lead to the detection of oscillation modes of amplitude as low as a few \cms\ after a few tens of hours of measurements. Such high-precision radial velocity measurements are of great importance for ground-based asteroseismic observations.

The F5 IV-V star Procyon A ($\alpha$\,CMi, HR\, 2943, HD\,61421) was chosen as the first target for an asteroseismic program with the new spectrometer \sophie\ at the 193-cm telescope of the Observatoire de Haute-Provence (OHP). The run, conducted in January 2007, was part of a large network campaign dedicated to this star. This campaign was proposed in order to improve the seismic analysis concerning this star and the modeling of its interior structure. This paper reports only the single-site observations with \sophie. It presents the performance of \sophie\ for asteroseismology, and shows that the high signal-to-noise ratio provided by this new instrument makes it possible to answer the questions raised by previous works that gave disparate values for the large spacing, e.g.
\cite{1998A&A...340..457M},
\cite{1999A&A...351..993M},
\cite{2004A&A...418..295M},
\cite{2005A&A...429L..17C},
\cite{2004A&A...422..247E},
\cite{2007A&A...464.1059L}.
The first ground-based observations reported different large spacings, in the range 70-80~$\mu$Hz (\cite{1986A&A...164..383G}; \cite{1991ApJ...368..599B}). Previous space-borne photometric observations with MOST were unable to detect solar-like oscillations in Procyon, or detected them very marginally (\cite{2005A&A...444L...5R}). Finally, Procyon is a too bright target for CoRoT (\cite{2002ESASP.485...17B}).

As shown by the long list of observations, Procyon is one of the most frequently studied asteroseismic targets. As a member of a double system, its fundamental parameters are now precisely determined. This is an advantage for a precise modeling based on asteroseismic constraints (see e.g. \cite{2006A&A...460..759P}). However, even if its bright magnitude ($m\ind{V}=0.4$) makes it an easy target for spectrometric observations, its asteroseismic characteristics are not so easily analyzed, for several reasons. First, the
Procyon p-modes frequencies lie in the frequency range [0.5, 1.5~mHz] where stellar
noise may be important. Second, both observations and modeling show that the large spacing is about 4.6 times greater than the 11.6-$\mu$Hz day alias. Also, large and  small spacing values provide a frequency difference between $\ell$=0 and $\ell$=1 modes very close to 2 $\times$ 11.6~$\mu$Hz. Finally, with mode linewidth as high as 2.5 $\mu$Hz according to \cite{1999A&A...351..582H}, the short lifetime of the pressure modes may hamper their detection.

Observations are presented in Sect.~\ref{observations}, with an introduction of the characteristics and performance of \sophie\ for asteroseismology.
The seismic analysis, with the identification of the large frequency spacing, is exposed in Sect.~\ref{signature}.
Asymptotic parameters are extracted from the Fourier spectrum and individual eigenfrequencies are identified. This identification of $\ell$ = 0 and 1 modes remains intricate in the day alias pattern. The discussion in Sect.~\ref{discussion} proposes a comparison with previous works dedicated to Procyon: this single-site run with an efficient instrument allows us to analyze the discrepancies reported in previous observations. Section~\ref{conclusion} is devoted to conclusions.


\section{Observations\label{observations}}

\begin{figure}
\centering
\includegraphics[width=8.5cm]{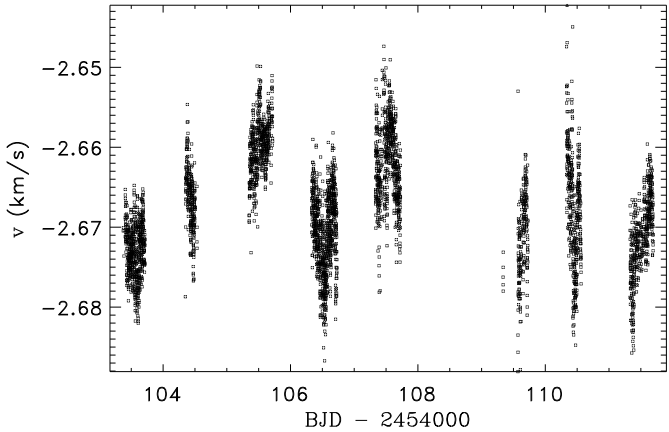}
\caption{Time series of the radial velocity of \cible\ measured with the pipeline reduction of \sophie\ (unfiltered data).
\label{timeseries}}
\end{figure}

\begin{figure}
\centering
\includegraphics[width=8.5cm]{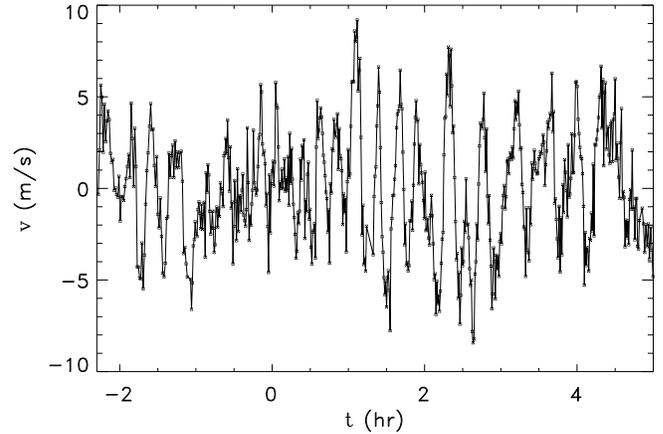}
\caption{Zoom of the radial velocity curve (1st night). Oscillations with a period about 17 min are clearly visible.}
\label{vitesse1}
\end{figure}

\subsection{The new spectrometer SOPHIE}

The observation of Procyon was carried out in January 2007 with the new spectrometer
\sophie\ based on the 1.93-m telescope of Haute Provence
Observatory (\cite{2006tafp.conf..319B}). The instrument, greatly motivated by the performances and success of the \harps\ spectrometer (\cite{2004A&A...423..385P}), is the successor of the \elodie\  spectrometer. One of its main properties consists in the optical dispersive components; they are installed in a constant-pressure tank, in order to avoid any drift due to atmospheric pressure variations.
Asteroseismic observations can be conducted in the high-resolution mode ($R=70\,000$) achieved with simultaneous Thorium-Argon lamp exposures. The spectrum, recorded on 39 spectral orders, ranges from 387 to 694~nm. The \sophie\ instrument was opened for observation in November 2006. In this first observation period, the instrument was still in its optimization phase, and offered to users who were informed of this status.

\subsection{Data reduction}

The spectra recorded with \sophie\ are extracted and automatically reduced in real-time,
using a data reduction software (DRS) adapted from the \harps\ DRS. Wavelength calibration is provided by the ThAr spectrum.
Radial velocities are obtained from weighted cross-correlation with a
numerical mask constructed from the Sun spectrum atlas.
The simultaneous ThAr spectrum gives the instrumental drift, using the optimum weight procedure of \cite{2001A&A...374..733B} based on the method proposed by \cite{1985Ap&SS.110..211C}.

The analysis of the RV achieved separately for each order showed that the last 4 red spectral orders, in the spectral range [642, 694~nm], present an abnormal behavior. The artefact comes from the pollution of the spectra due to the strong Argon lines in this spectral range. The optical filter designed to cut these Argon lines for the \sophie\ spectral format was not yet installed at the time of the Procyon observation, thus the red part of the spectra was strongly polluted. The cross-correlation function computed without these last 4 spectral orders resulted in a significant improvement of the RV curve.

In the Procyon observations, the spectrometer drift has proved to be systematically lower than 30\ms\ during a full night, and most often lower than 10\ms. However, a rapidly varying term, about 4\ms, was observed, related to the turbulence inside the insulation box generated by the thermal control air flow.
This effect increases with decreasing exposure times, and is no more significant for exposure time longer than 3~min. Fortunately, this turbulence affects both stellar and ThAr beams in the same way, therefore its influence is quite well corrected.
Recent improvement of the insulation box of \sophie\ has permitted the reduction of this effect of turbulence to below 1\ms\ for short exposure and to reduce the slow drift over the night below 10\ms.

\begin{table}
\caption{Journal of radial velocity measurements, with the minimum and maximum spectrum SNR. $\Delta T$ represents the length of observation each night.
The dispersion $\sigma\ind{RMS}$ is directly measured in the time series for each night. This term is dominated
by the seismic signal and does not reflect the precision of individual points. The dispersion $\sigma\ind{RV}$ is derived from the high frequency noise recorded in the power spectrum of each night. \label{journal}}
\begin{center}
\begin{tabular}{ccccccc} \hline Date & Number of & $\Delta T$ & SNR & $\sigma\ind{RMS}$ &  $\sigma\ind{RV}$  \\
Jan. 07 & spectra & (hr) &  & (\ms) & (\ms)\\
\hline
02 & 596  & 8.18 & 139-560  & 3.3  & 1.4 \\ 
03 & 255  & 4.41 & 110-735  & 3.9  & 1.9 \\ 
04 & 507  & 8.83 & 132-781  & 3.5  & 1.4 \\ 
05 & 849  & 9.50 & 163-661  & 4.8  & 2.6 \\ 
06 & 629  & 9.43 & 101-876  & 4.9  & 2.7 \\ 
08 & 220  & 3.70 & 107-736  & 5.3  & 3.1 \\ 
09 & 405  & 5.50 & 130-854  & 7.0  & 3.4 \\ 
10 & 466  & 8.70 & 113-1090 & 4.8  & 1.8 \\ 
\hline
\end{tabular}
\end{center}
\end{table}

\subsection{Time series properties}

About 52.4 hours of observation were recorded between January 2007, 2 and 11,
corresponding to 3927 individual measurements. Due to the bright magnitude of Procyon, the exposure time had to be carefully adapted to the seeing conditions, and varied from 7 to 30~s. Taking into account the CCD readout time (28~s), the sampling time of the signal varied between 35 and 58 s, for a Nyquist frequency about 10.2~mHz, much higher than the observed frequency range of the p-modes. With 1 night lost because of bad weather, 3 nights partly cloudy, and 5 with a mean duration about 9 hr, for 8 hr effective observation at high signal-to-noise ratio data, the final duty cycle was about 26\%. A journal of the observations is given in Table~\ref{journal}.

The radial velocity time series is presented in Fig.~\ref{timeseries} without any filtering
process. 
The dispersion of each individual night time series (Table~\ref{journal}) is strongly dominated by the acoustic modes with a period around 17~min, as shown in Fig.~\ref{vitesse1}.
This figure should be directly compared to the Fig.~1 of Bouchy et al. (2004), which displays the RV of Procyon obtained with the \harps\ spectrometer.


\subsection{Performance}

Under typical conditions, the signal-to-noise ratio per pixel (namely 0.025 \AA) at 550 nm was in the range 500-600. For Procyon, such a SNR corresponds to a photon-noise limited performance of about 0.45\ms. Compared to \elodie, the efficiency of \sophie\ is about 8 times better (2.3 mag). The gain in spectral resolution power is about 1.7. \cite{1999A&A...351..993M} obtained a photon-noise uncertainty of 0.8\ms\ for a 40-s exposure with \elodie\ on Procyon. Scaled to the same exposure time, the gain provided by \sophie\ for photon-noise limited measurement is about a factor of 3.

The standard deviation of the time series signal, including both seismic signal and noise, varies from 3.3 to 7\ms, depending on the seeing and sky transparency. 
The high frequency noise is about 5 cm\,s$^{-1}$, much less than that recorded in previous observations. \cite{2007A&A...464.1059L} report 30 cm\,s$^{-1}$ with \textsc{sarg},  \cite{2004A&A...422..247E} 11 cm\,s$^{-1}$ with \textsc{coralie},
\cite{2004A&A...418..295M} 8.6 cm\,s$^{-1}$ with \elodie. We note that the real gain  compared to \elodie\ is lower than 3. Photon-noise limited performance is not achieved, as expected for such a target as bright as Procyon, as discussed below.

\subsection{Power spectrum analysis}

\begin{figure}
\centering
\includegraphics[width=8.5cm]{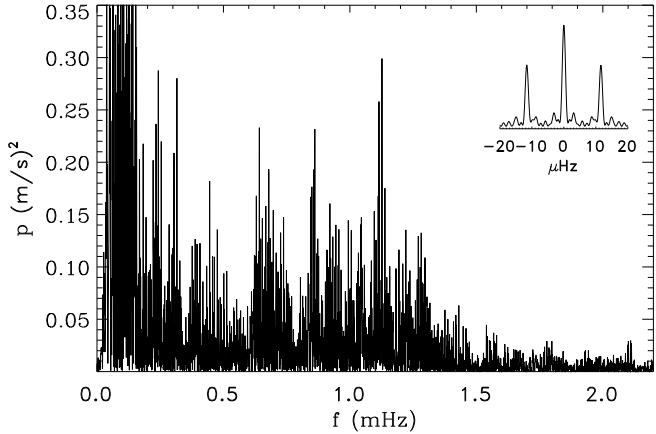}
\caption{Fourier spectrum of Procyon, and inset of the window function. The time series used for this spectrum excludes the noisiest values.
\label{spectrum}}
\end{figure}

In order to reduce the low frequency instrumental noise, the noisiest data (above 5 $\sigma$) were first excluded from the time series. Then, a 2-order polynomial fit was calculated for each night, and subtracted. This correction accounts for slowly varying drift residuals not corrected by the simultaneous ThAr measurement: atmospheric effects, or stellar signature, such as activity. The resulting Lomb-Scargle periodogram of the filtered time series is
shown in Fig.~\ref{spectrum}. Normalization of the spectrum ensures that a sinusoid with a rms value of  $v\ind{rms}$ will show a peak with power $v\ind{rms}^2$. The spectrum exhibits a series of peaks between 0.5 and 1.5 mHz, modulated by a broad envelope, which is the typical signature of solar-like oscillations. This signature appears in the power spectrum of each individual night.

The mean white noise level computed in the power spectrum in the range 4-6 mHz is about $\sigma\ind{HF}^2$=0.0026\m2s2, or 1.40\ms\,mHz$^{-1/2}$.
The corresponding velocity noise thus corresponds to 2.2\ms, according to the normalization $\sigma\ind{RV}=\sqrt{N/2}\ \sigma\ind{HF}$. 
The RV dispersions $\sigma\ind{RV}$ computed for each night show important variations (Table~\ref{journal}). These variations cannot be due to the instrument, since it was operated in the same conditions throughout the run. However, strong seeing changes  occurred, and it appears that good seeing is correlated with noisy measurements. We therefore strongly suspect guiding noise due to imperfect centering on the optical fiber input to explain most of the additional noise. The acceptance of \sophie\ fiber is 3 arcsec on the sky, in adequation
with the 2.5 arcsec median seeing at OHP. When the seeing is significantly smaller than 2 arcsec, short time exposures preclude a good average position of the stellar beam on the fiber input.
This effect was also a limitation with \harps\ observations of $\alpha$ Cen A, as
discussed and described by \cite{2007A&A...470..295B}. Dimmer targets requiring longer exposures will not be affected by this artefact.

\section{The seismic signature\label{signature}}

\subsection{\'Echelle diagram}

The velocity time series was analyzed according to the method based on the false alarm probability for detecting peaks embedded in noise described in \cite{2004A&A...428.1039A}. This method excludes any a-priori information for the peak selection.
The H0 hypothesis corresponds to a pure white noise contribution. Application of the test makes it clear that this hypothesis is rejected, with many peaks much greater than 10\,$\sigma$ in the power spectrum in the range [0.5 - 1.5\,mHz]. The next step of the peak identification relies on the assumption that the global description of low-degree p-modes derives from the asymptotic theory: modes are  expected to obey the asymptotic comb-like pattern, which permits an \'echelle diagram analysis.

\begin{figure}
\centering
\includegraphics[width=8.5cm]{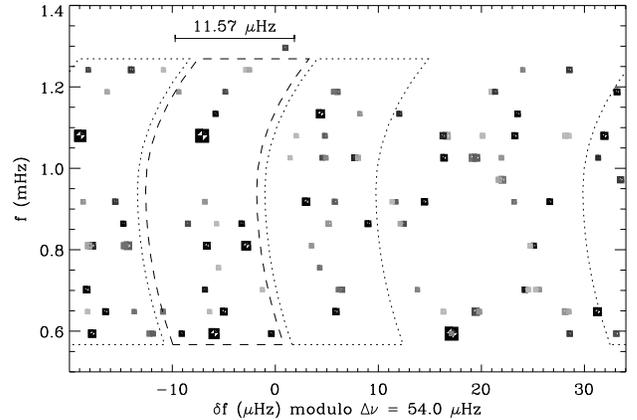}
\caption{\'Echelle diagram constructed with the peaks selected with the lowest false alarm probability (a threshold of 10\% has been chosen for a large selection). The size of the symbols represents the peak amplitudes. For clarity, the x-axis presents a negative offset. Despite the omnipresent signature of the day aliases, ridges can be detected. The region surrounded with a dashed contour  corresponds to the most prominent ridge; its aliases are delimited with dotted lines. 
\label{detect}}
\end{figure}

Except for the regularity detected around 54~$\mu$Hz, the \'echelle diagram is highly dominated by the day aliases, as expected for single-site measurement (Fig.~\ref{detect}). However, it clearly shows regions with much less signal, delimiting ridges. This procedure allows us to locate the ridge with maximum energy, around $-6\,\mu$Hz (modulo 54~$\mu$Hz).
Performing a collapsogram is a classical way to assess the presence of ridges. However, since the \'echelle diagram of Procyon shows significant departure from vertical alignments, the collapsogram cannot provide clear information.
Therefore, rescaling has been processed before its construction. In practice, each main peak identified in the main ridge is first recentered at the null frequency of the collapsogram. In a second step, each frequency interval between two consecutive peaks, whose extent corresponds to the large separation between these peaks, is rescaled to a fixed value of the large separation $(\Delta\nu\ind{S})$. This rescaling is obtained by expanding or shrinking the frequency axis in order to fit to the fixed large separation, then by interpolating the portion of the spectrum over fixed frequency values between 0 and $\Delta\nu\ind{S}$.
Finally, the addition of all rescaled portions of the spectrum provides the rescaled collapsogram (Fig.~\ref{collapsogram}).
This procedure allows us to correct the collapsogram for systematic discrepancies to a vertically aligned pattern, but its efficiency is reduced by the expected scattering due to the limited lifetimes of the modes. This scattering appears much less pronounced than the modulation in the \'echelle diagram, so that the rescaling process is efficient.
Hence, the collapsogram allows us to verify the location of the ridge with maximum energy. Aliases located at $\pm$11.6~$\mu$Hz of this ridge (namely at about $-18$ and 5\,$\mu$Hz in Fig.~\ref{detect}) were not considered in the mode analysis, and only the major peaks of the whole pattern were kept for a simplified \'echelle diagram (Fig.~\ref{echelle}).

\subsection{Asymptotic parameters\label{asymp}}

\begin{figure}
\centering
\includegraphics[width=8.5cm]{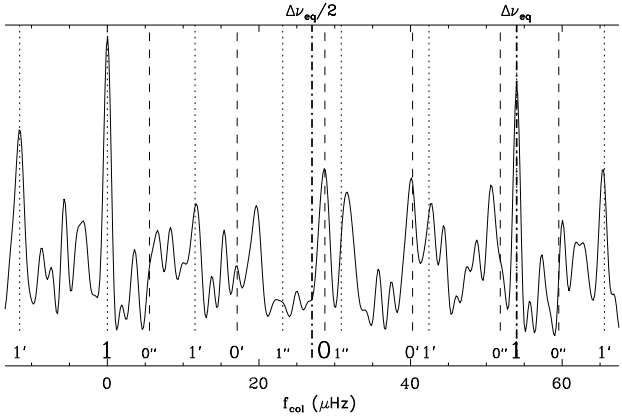}
\caption{Rescaled collapsogram of the \'echelle diagram of Fig.~\ref{detect}, based on the $\ell=1$ ridge and rescaled on a fixed value of the large frequency spacing. The signature of $\ell$=0 mode is clearly identified, slightly greater than half the large spacing (dot-dashed line), but there is no clear detection of $\ell=2$ modes. Most of the peaks are related to the $\ell$=0 (dashed lines) and $\ell$=1 (dotted lines) modes, or to their aliases at $\pm$11.6\,$\mu$Hz (') or $\pm23.2\,\mu$Hz (''). The alias at $-23.2\,\mu$Hz of the $\ell$=1 ridge appears to be boosted in the vicinity of the $\ell$=0 ridge.
\label{collapsogram}}
\end{figure}

\begin{figure}
\centering
\includegraphics[width=8.5cm]{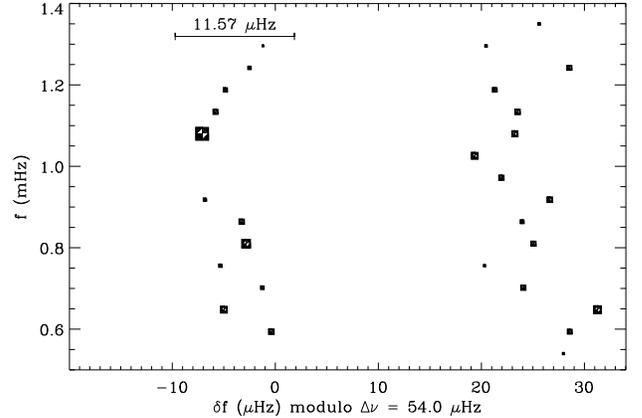}
\caption{
\'Echelle diagram, selecting only the major peaks detected in Fig.~\ref{detect}, and excluding the peaks present in the neighboring alias of the ridge around $-6\,\mu$Hz (modulo 54~$\mu$Hz), identified to $\ell$=1 modes. The other ridge includes $\ell$=0 and, maybe, 2 modes. The highest confidence levels are reached for the peaks identified above 1~mHz, which exhibit large spacings about 55\,$\mu$Hz.
\label{echelle}}
\end{figure}

We proposed a fit of the selected frequencies according to the following terms:
$$
\nu_{n,\ell} = \nu_{n_0} + n_\ell \ \Delta\nu
- \ell (\ell + 1) \ D_0
+ {n_\ell^2 \over 2}\ \courb
$$
with $n_0$ the radial order of a suspected $\ell$=0 peak at frequency $\nu_{n_0}$ and
$$n_\ell = n -n_0 + {\ell\over 2}
$$
The frequency $D_0$ measures the small spacing. The introduction of the parameter $\courb$ is actually necessary to take into account the local curvature in the \'echelle diagram. However, it relates only a linear increase of the large spacing with the radial order $n$ ($\Delta\nu (n) = \Delta\nu (n_0) + \courb\, n_\ell$).
The collapsogram (Fig.~\ref{collapsogram}), corrected for the variation of the large spacing with frequency, allows us to measure $D_0$. Simultaneously, the necessarily positive value of $D_0$ gives the identification of the $\ell$=0 and $\ell$=1 ridges.
It also demonstrates the invasive influence of the day aliases: the frequency differences  $\nu_{n,1} - \nu_{n,0}$ and $\nu_{n,2} - \nu_{n,1}$ are both close to 2$\times$11.6~$\mu$Hz.

\subsection{Day alias and frequency spacings}

Simulations were performed, following the procedure in \cite{1990ApJ...364..699A}, in order to analyze this coincidence precisely.
We have calculated different synthetic spectra, assuming an asymptotic pattern, with $\Delta\nu$ and $\courb$ given by the fit of the data, but with a varying $D_0$
frequency. Mode linewidths, about 2.5\,$\mu$Hz, were derived for a Procyon model from \cite{1999A&A...351..582H}.
Simulations were run with the same window function and SNR as the observations. They also considered the visibility of the modes derived from the value of inclination (31.1$\pm$0.6$^\circ$) reported by \cite{2000AJ....119.2428G}.

Independently of the $D_0$ value, the simulations first show the prominent role of the $\ell$=1 ridge with its aliases at $\pm$11.6~$\mu$Hz, when $\ell$=0 modes are strongly perturbed by the vicinity of $\ell$=2 modes. Second, they demonstrate the possible interference between the day alias of the  $\ell$=0 and 1 modes located between the $\ell$=0 and 1 ridges.

The simulations indicate that the value of $6D_0$ ($\simeq\deuxzero=\nu_{n,0}-\nu_{n-1,2}$) must be close to  7.6\,$\mu$Hz. This value corresponds to $\Delta\nu - 4\times 11.6\,\mu$Hz, which means the spectrum is organized by the day aliases. In fact, values of $D_0$ significantly different from this case would yield an interruption of the regular spacing due to the day aliases, which is not observed in Fig.~\ref{detect}.

\begin{figure}
\centering
\includegraphics[width=8.5cm]{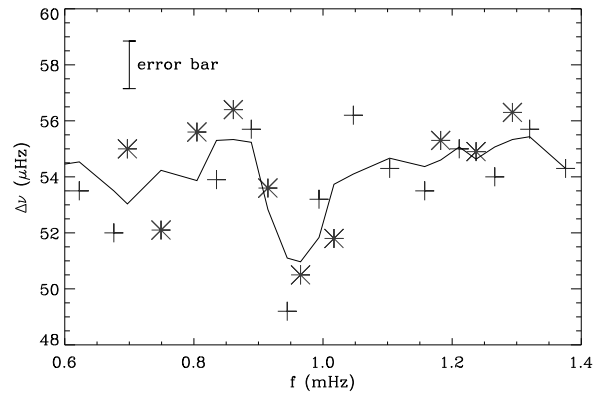}
\caption{
Large spacing as a function of frequency, for the degrees 0 ($+$), 1 ($\ast$) (after disentangling of $\ell$=0 and 2 modes), and global fit (solid line) corresponding to the boxcar average smoothed over 3 consecutive values.
Error bars, estimated according to \cite{1992ApJ...387..712L}, are about 1.6 $\mu$Hz.
As in \cite{2004A&A...418..295M}, a lower value of the large spacing is observed in the frequency range [0.9, 1.0\,mHz]. Outside this range, the large spacing increases with frequency.
\label{fig7}}
\end{figure}

Finally, after identification of the major peaks with the largest amplitudes and confidence levels, 
we can derive the local values of the large spacing (Fig.~\ref{fig7}), varying with frequency. More complicated variations are also suspected, as observed in \cite{2004A&A...418..295M} and modeled in \cite{2006A&A...460..759P}.

Due to the short mode lifetimes mentioned by \cite{1999A&A...351..582H} or derived by \cite{2007A&A...464.1059L}, and due to the short observation duration, the scatter on the measured large separations is as high as 1.5\,$\mu$Hz. Defining a single value is not possible, since the large spacing increases at a rate of about $\diff \Delta\nu/\diff n \simeq 0.2\,\mu$Hz, except in the range [0.9, 1.0 mHz].
The small spacing derived from the collapsogram is about $1.0\,\mu$Hz. Since the collapsogram has been constructed after a recentering and rescaling process, deriving the exact value of $D_0$ cannot be achieved precisely.

\subsection{Amplitudes}

The Monte-Carlo simulations were run for an estimate of the modes amplitudes in the [0.6 - 1.4 mHz] frequency range, assuming the contribution of this synthetic signal with a global Gaussian envelope plus white noise. The best fit corresponds to an envelope centered at 0.95\,mHz and with a 0.9\, mHz FWHM, and the best agreement between the synthetic and observed spectra occurs for a signal maximum amplitude about 45$\pm$8\, cm\,s$^{-1}$ rms. Uncertainties are mainly related to the assumptions on the parameters of the amplitude envelope.

This estimate is in full agreement with previous works; \cite{2004A&A...418..295M} and \cite{2007A&A...464.1059L} also report 45\,cm\,s$^{-1}$, although the simulations by Marti\'c et al. give the best agreement with the maximum mode amplitude of about 35\,cm\,s$^{-1}$.
We note that the simulations often show a 2-bump energy excess, certainly related to the combined variations with frequency of the mode linewidths (Houdek et al. 1999) and of the smooth envelope of the amplitudes. We remark that the use of the H1 hypothesis of \cite{2004A&A...428.1039A} provides the best detection probability with a bin about 4\,$\mu$Hz, consistent with short mode lifetimes. However, this hypothesis has been of little interest for determining the mode amplitudes, since its use supposes a precise determination of the background noise, which is hampered by the high mode density and by the many aliases.


\section{Discussion\label{discussion}}

Previous papers have reported different values of the large spacing,
around 56 or 54~$\mu$Hz:
\cite{2004A&A...422..247E} report 55.5\,$\mu$Hz,
\cite{2005A&A...429L..17C} 56$\pm 2$,
\cite{2007A&A...464.1059L} 55.9$\pm$0.08,
\cite{1998A&A...340..457M} 53$\pm$3,
\cite{2004A&A...418..295M} 53.6$\pm$0.5. All measurements were derived from mode identification, except \cite{1998A&A...340..457M} and \cite{2005A&A...429L..17C}, who report values derived from the comb analysis.
The \sophie\ measurements are clearly in favor of large spacings varying significantly with frequency (Fig.~\ref{fig7}). These variations were already reported in \cite{2004A&A...418..295M}.
An \'echelle diagram based on the mean value 54\,$\mu$Hz seems preferable than based on 56~$\mu$Hz; peaks are continuously detected, from 0.6 to 1.3~mHz, with only small departures from an asymptotic pattern. On the contrary, a uniform large spacing of 56~$\mu$Hz would imply important deviations from the comb-like pattern in the frequency range [0.9 - 1.0~mHz].

As in most previous observations, we observe an energy gap around 1.0 mHz. This recurrent gap is very plausibly related to a competition between the excitation efficiency and the mode lifetimes; less energy is transferred to low-frequency modes below 1.0 mHz, but their longer lifetimes yield larger amplitudes than in the [0.9-1.0~mHz] frequency range. The presence of this gap associated with the curvature in the oscillation pattern leads, in noisy data, to 2 possible solution sets, corresponding crudely to comb-like patterns constructed with a 54 or 56-$\mu$Hz frequency spacing. A wrong solution may appear when alignments are misplaced, based on a given ridge in the highest part of the spectrum, but aligned with one of its alias at a frequency lower than the energy gap, or vice-versa. In the \sophie\ data, the solution of a \emph{mean} large spacing value around 54\,$\mu$Hz is preferred (Fig.~\ref{detect}).
In the frequency ranges below 0.9\,mHz and above 1\,mHz, the increase of the large spacing is, however, clearly visible (Fig.~\ref{fig7}), showing that most of the large spacing values are in the range 54-56\,$\mu$Hz. 

We note that the eigenfrequencies we have found are compatible with those observed by \cite{2004A&A...422..247E}, or with  their aliases, but with different degree identifications.
The identification of the ridges we propose is consistent with that reported in  \cite{2004A&A...418..295M}, i.e. with the frequency agreeing within about 11.6\,$\mu$Hz, to take into account the possible confusion due to day aliases. However, the mode identifications reported in this paper are not compatible with those reported in \cite{2004A&A...418..295M}.
The discrepancy consists in an inversion between the $\ell$=1 and $\ell$=0 (or 2) modes. As indicated by the simulations reported in Sect.~\ref{asymp}, a definitive conclusion concerning this identification is not possible. Furthermore, a few mixed modes are certainly present (Marti\'c et al. 2004), which complicates the analysis. The data set in \cite{2004A&A...418..295M} is based on observations with a better duty cycle, whereas these new single-site data with the high SNR provided by \sophie\ allow a more direct mode identification based on statistical tests. Both observation data sets agree in identifying a significant deviation compared to a regular comb-like structure. Network data will definitely close the controversy.



\section{Conclusion\label{conclusion}}

The performance of the new spectrometer \sophie\ at OHP appears remarkable. For asteroseismology, it is improved by a factor of 3 compared to the performance reached by the previous spectrometer \elodie. Therefore, asteroseismic observations on targets down to the magnitude 6 can be achieved with \sophie. For Procyon, performance is not photon-noise but seeing limited, due to imperfect centering of the input fiber for short exposures. However, this drawback will not affect dimmer targets. Furthermore, a new Cassegrain fiber adapter with a high-frequency guiding system will contribute to reducing this artefact.

Previous observations of Procyon presented different values for the large spacing and contradictory mode identifications. In the \sophie\ data, however, about \nident\ $\ell$=0 or 1 peaks were identified in the whole spectrum, even in the frequency range around [0.9, 1.0~mHz] corresponding to lower amplitudes, and where rapid variations of the large frequency spacing occur. In fact, the sensitivity reached with \sophie\ allows us to reconcile previous observations: the \'echelle diagram analysis has permitted the identification of variations of the large spacing with frequency.
In the interval [0.9, 1.0~mHz], low values are reached, down to 52\,$\mu$Hz. In the frequency ranges below 0.9\,mHz and above 1.0\,mHz, the large spacing increases at a rate of about $\diff \Delta\nu/\diff n = 0.2\pm0.1\,\mu$Hz, this value being independent on the degree identification. In the frequency range above 1\,mHz, where peaks are identified with the highest confidence level, large spacings reach values of about 55\,$\mu$Hz (increasing from 54 to 56\,$\mu$Hz). This work also shows that the identification of the mode degrees is extremely difficult with single-site measurement, since the values of the large and small spacing ($D_0\simeq 1\,\mu$Hz) yields strong interferences between the aliases at $\pm$23.2\,$\mu$Hz of $\ell$=1 modes with $\ell$=2 and $\ell$=0 modes.

These results show that the performance of the new spectrometer \sophie\ is among the best for asteroseismology. The results show promise for the outcome of the combined data resulting from the network campaign.

\begin{acknowledgements}
FB thanks all the staff of the Observatoire de Haute Provence for their work in building and operating the spectrometer \sophie. We are grateful to C. Carol, H. Le Coroller, J.P. Meunier, J. Topenas for the on-site support of the observations. We thank the anonymous  referee for his useful comments.

\end{acknowledgements}


\begin{thebibliography}{}

\bibitem[Anderson et al. (1990)]{1990ApJ...364..699A} Anderson, E.~R.,
Duvall, T.~L., Jr., \& Jefferies, S.~M.\ 1990, \apj, 364, 699
\bibitem[Appourchaux (2004)]{2004A&A...428.1039A} Appourchaux, T.\ 2004, \aap, 428, 1039

\bibitem[Baglin et al. 2002]{2002ESASP.485...17B} Baglin, A., Auvergne,
M., Barge, P., Buey, J.-T., Catala, C., Michel, E., Weiss, W., \& COROT
Team 2002, Stellar Structure and Habitable Planet Finding, 485, 17



\bibitem[Bazot et al. (2007)]{2007A&A...470..295B} Bazot, M., Bouchy, F.,
Kjeldsen, H., Charpinet, S., Laymand, M., \& Vauclair, S.\ 2007, \aap, 470,
295

\bibitem[Bedding \& Kjeldsen 2006]{2006ESASP.624E..25B} Bedding, T., \&
Kjeldsen, H.\ 2006, ESA Special Publication, 624,

\bibitem[Brown et al. 1991]{1991ApJ...368..599B} Brown, T.~M., Gilliland,
R.~L., Noyes, R.~W., \& Ramsey, L.~W.\ 1991, \apj, 368, 599

\bibitem[Bouchy et al. (2001)]{2001A&A...374..733B} Bouchy, F., Pepe, F., \&
Queloz, D.\ 2001, \aap, 374, 733

\bibitem[Bouchy et al. 2004]{bouchy04}
Bouchy F., et al. 2004, Nature, 432, 7015

\bibitem[Bouchy \& The Sophie Team 2006]{2006tafp.conf..319B} Bouchy, F.,
\& The Sophie Team 2006, Tenth Anniversary of 51 Peg-b: Status of and
prospects for hot Jupiter studies, 319

\bibitem[Claudi et al. (2005)]{2005A&A...429L..17C} Claudi, R.~U., et al.\
2005, \aap, 429, L17

\bibitem[Connes (1985)]{1985Ap&SS.110..211C} Connes, P.\ 1985, \apss, 110,211

\bibitem[Eggenberger et al. (2004)]{2004A&A...422..247E} Eggenberger, P.,
Carrier, F., Bouchy, F., \& Blecha, A.\ 2004, \aap, 422, 247


\bibitem[Gelly et al. 1986]{1986A&A...164..383G} Gelly, B., Gree, G., \&
Fossat, E.\ 1986, \aap, 164, 383

\bibitem[Girard et al. (2000)]{2000AJ....119.2428G} Girard, T.~M., et al.\
2000, \aj, 119, 2428

\bibitem[Houdek et al. (1999)]{1999A&A...351..582H} Houdek, G., Balmforth, N.~J., Christensen-Dalsgaard, J., \& Gough, D.~O.\ 1999, \aap, 351, 582

\bibitem[Leccia et al. (2007)]{2007A&A...464.1059L} Leccia, S., Kjeldsen,
H., Bonanno, A., Claudi, R.~U., Ventura, R., \& Patern{\`o}, L.\ 2007,
\aap, 464, 1059

\bibitem[Libbrecht (1992)]{1992ApJ...387..712L} Libbrecht, K.~G.\ 1992,
\apj, 387, 712

\bibitem[Marti{\'c} et al. (1999)]{1999A&A...351..993M} Marti{\'c}, M., et
al.\ 1999, \aap, 351, 993

\bibitem[Marti{\'c} et al. (2004)]{2004A&A...418..295M} Marti{\'c}, M.,
Lebrun, J.-C., Appourchaux, T., \& Korzennik, S.~G.\ 2004, \aap, 418, 295


\bibitem[Mosser et al. (1998)]{1998A&A...340..457M} Mosser, B., Maillard,
J.~P., M\'ekarnia, D., \& Gay, J.\ 1998, \aap, 340, 457

\bibitem[Pepe et al. 2004]{2004A&A...423..385P} Pepe, F., et al.\ 2004,
\aap, 423, 385

\bibitem[Provost et al. 2006]{2006A&A...460..759P} Provost, J.,
Berthomieu, G., Marti{\'c}, M., \& Morel, P.\ 2006, \aap, 460, 759

\bibitem[R{\'e}gulo \& Roca Cort{\'e}s 2005]{2005A&A...444L...5R}
R{\'e}gulo, C., \& Roca Cort{\'e}s, T.\ 2005, \aap, 444, L5




\end{thebibliography}
\end{document}